\documentclass{article}
\usepackage{templates/spconf,amsmath,amsfonts,epsfig,hyperref}
\usepackage[table,xcdraw]{xcolor}
\usepackage{booktabs}       
\usepackage{subfig}
\let\OLDthebibliography\thebibliography
\renewcommand\thebibliography[1]{
  \OLDthebibliography{#1}
  \setlength{\parskip}{0pt}
  \setlength{\itemsep}{0pt plus 0.3ex}
}
\usepackage{paralist}
\usepackage{enumitem}
\usepackage{amssymb}
\usepackage{amsthm}
\usepackage{mathrsfs}
\usepackage{indentfirst}
\usepackage{multirow}
\begin{document}\sloppy
\title{A-ESRGAN: Training Real-World Blind Super-Resolution with Attention U-Net Discriminators}
%

\newcommand{\repository}{\url{https://github.com/aesrgan/A-ESRGAN}}

\name{
Zihao Wei$^{1,2}$,  Yidong Huang$^{1,2}$, Yuang Chen $^{1,2}$, Chenhao Zheng$^{1,2}$, Jinnan Gao $^{2}$}

\address{ $1$ Department of Computer Science Engineering, University of Michigan, Ann Arbor, USA\\
$2$ UM-SJTU Joint Institute, Shanghai Jiao Tong University, Shanghai, China\\
\{zihaowei, owenhji, cyaa, neymar\}@umich.edu; \{gjn0310\}@sjtu.edu.cn \\
\repository
}

\maketitle
\begin{abstract}
Blind image super-resolution(SR) is a long-standing task in CV that aims to restore low-resolution(LR) images suffering from unknown and complex distortions. Recent work has largely focused on adopting more complicated degradation models to emulate real-world degradations. The resulting models have made breakthroughs in perceptual loss and yield perceptually convincing results. However, the limitation brought by current generative adversarial network(GAN) structures is still significant: treating pixels equally leads to the ignorance of the image’s structural features, and results in performance drawbacks such as twisted lines \cite{Real-ESRGAN} and background over-sharpening or blurring.  In this paper, we present A-ESRGAN, a GAN model for blind SR tasks featuring an attention U-Net based, multi-scale discriminator that can be seamlessly integrated with other generators. To our knowledge, this is the first work to introduce attention U-Net structure as the discriminator of GAN to solve blind SR problems. And the paper also gives an interpretation for the mechanism behind multi-scale attention U-Net that brings performance breakthrough to the model. Through comparison experiments with prior works, our model presents state-of-the-art level performance on the non-reference natural image quality evaluator(NIQE) \cite{NIQE} metric. And our ablation studies have shown that with our discriminator, the RRDB \cite{ESRGAN} based generator can leverage the structural features of an image in multiple scales, and consequently yields more perceptually realistic high-resolution(HR) images compared to prior works. 

\end{abstract}
\section{Introduction and Motivation}

Image super-resolution (SR) is a low-level computer vision problem aiming to reconstruct a high-resolution(HR) image from a distorted low-resolution(LR) image. Blind super-resolution, specifically, refers to the idea of restoring LR images suffering from unknown and complex degradation, as opposed to the traditional assumption of ideal bicubic degradation.

In recent years, the main methods of this field have been dominated by deep learning. Specifically, the trend started from SRCNN \cite{SRCNN}, a convolutional neural network model which achieved notable performance. However, while these methods are able to generate images with high especially in Peak Signal-to-Noise Ratio (PSNR) value, they tend to output over-smoothed results which lack high-frequency details \cite{ESRGAN}. Therefore, scholars proposed to use generative adversarial networks(GANs) to solve image super-resolution challenges. A super-resolution GAN composes of a generator network and a discriminator network, in which the generator takes LR images as input and aims to generate images as similar to the original high-resolution image as possible, while the discriminator tries to distinguish between "fake" images generated by the generator and real high-resolution images. 

By the competition of generator and discriminator, the networks are encouraged to favor solutions that look more like natural images. The state-of-the-art methods using generative adversarial network includes
ESRGAN,Real-ESRGAN and BSRGAN\cite{ESRGAN,Real-ESRGAN,BSRGAN}.

Recent work in super-resolution GAN has largely focused on simulating a more complex and realistic degradation process \cite{Real-ESRGAN} or building a better generator \cite{ESRGAN}, with little work trying to improve the performance of the discriminator. However, the importance of a discriminator can not be ignored since it provides the generator the direction to generate better images, much like a loss function. In this work, we construct a new discriminator network structure: \textbf{Multi-scale Attention U-Net Discriminator} and incorporate it with the existing RRDB based generator \cite{ESRGAN} to form our GAN model A-ESRGAN. Our model shows superiority over the state-of-the-art real-ESRGAN model in sharpness and details (see \autoref{fig:AttentionUnetAblationDistortion}). This result owes to the combination of attention mechanism and U-Net Structure in our proposed discriminator.  U-Net Structure in discriminator can provide per-pixel feedback to the generator\cite{Schonfeld_2020_CVPR}, which can help the generator to generate more detailed features, such as texture or brushstroke. Meanwhile, the attention layer can not only distinguish the outline of the subject area so as to maintain the global coherence but strengthen the lines and edges of the image to avoid the blurring effect (this is demonstrated in the attention map analysis section in our paper). Therefore, the combination of U-Net and Attention is very promising. Besides, in order to increase the perception field of our discriminator, We use 2 attention U-Net discriminators that have an identical network structure but operate at different image scales as our final discriminator, which is called multi-scale discriminator. Extensive experiments show that our model outperforms most existing GAN models both in quantitative NIQE performance metric and qualitative image perceptual feelings.

In summary, the contributions of our work are:
\begin{itemize}
    \item We propose a new multi-scale attention U-Net discriminator network. To the best of our knowledge, this is the first work to adopt attention U-Net structure as a discriminator in the field of generative adversarial network. This modular discriminator structure can be easily ported to future work.
    \item We incorporate our designed discriminator with the existing RRDB based generator to form our generative adversarial network model A-ESRGAN. Experiments show that our model outperforms most state-of-the-art models in image super-resolution tasks.
    \item Through detailed analysis and visualization about different layers of our network, we provide convincing reasoning about why multi-scale attention U-Net discriminator works better than existing discriminators in image super-resolution tasks.
\end{itemize}
\section{Related Work}
Since the paper focuses on designing an improved multi-scale discriminator by leveraging attention U-Net to train a GAN model for blind SR tasks, we will give a brief overview on related GANs-based blind SR works.

\textbf{GANs-based Blind SR Methods} Before GAN framework is applied, deep convolutional neural networks(CNNs) are widely adopted\cite{SRCNN, SRCNN-cont0, SRCNN-cont1} in the field of blind image SR tasks. Owing to CNN's strong modeling power, these methods have achieved impressive  PSNR performance. However, because these PSNR-oriented methods use pixel-wise defined losses such as MSE\cite{SRCNN}, the model tends to find the pixel-wise average of multiple possible solutions, which generally leads to overly-smoothed results and absence of high-frequency details like image textures \cite{ESRGAN}. Some scholars proposed GANs-based approaches \cite{SRGAN, SRGAN-cont0, SFT-GAN} to address the aforementioned problem, because GANs have been proven competitive in learning a mapping between manifolds and can therefore improve the reconstructed local textures \cite{manifold}. Recent state-of-the-art works have raised a perceptual-driven perspective to improve GANs by better modeling the perceptual loss between images\cite{SRGAN, ESRGAN}. The ESRGAN\cite{ESRGAN}, as a representative work, proposed a practical perceptual loss function as well as a residual-in-residual block(RRDB) generator network, and produces synthesized HR images with convincing visual quality. Another perspective is to solve the intrinsic problem of blind SR that the LR images used for training are synthesized from HR images in the dataset. Most existing methods are based on bicubic downsampling \cite{SRCNN, lai2017deep, sajjadi2017enhancenet} and traditional degradations \cite{michaeli2013nonparametric, zhang2015revisiting, zhang2021plug, zhang2018image}, while real-world degradations are far more complicated. To produce more photo-realistic results, the real-ESRGAN \cite{Real-ESRGAN} proposed a practical high-order degradation model and achieved visually impressive results as well as state-of-the-art NIQE \cite{NIQE} performance. Our work is based on the degradation model and RRDBN generator of Real-ESRGAN, and we propose a novel and transportable discriminator model named attention U-Net to remedy the limitation of current GANs architectures.

\par\noindent\textbf{Discriminator Models} Some remarkable attempts have been made to improve the discriminator model\cite{Wang_2018_CVPR, Schonfeld_2020_CVPR, park2018srfeat}. To synthesize photo-realistic HR images, two major challenges are presented: the discriminator needs a large receptive field to differentiate the synthesized image and the ground truth(GT), requiring either deep network or large convolution kernel \cite{Wang_2018_CVPR}. Besides, it's difficult for one discriminator to give precise feedback on both global and local features, leading to possible incoherence in the synthesized image such as twisted textures on a building wall \cite{Real-ESRGAN}. Wang et al. \cite{Wang_2018_CVPR}
proposed a novel multiple discriminator architecture to resolve these two issues. One discriminator accepts down-sampled synthesized images as input and has a larger receptive field with fewer parameters, and it's responsible to grasp the global view. The other discriminator takes the full synthesized image as input to learn the details. Another pioneer work \cite{Schonfeld_2020_CVPR} introduces U-Net based discriminator architecture into GANs-based blind SR tasks. The U-Net discriminator model can provide per-pixel feedback to the generator while maintaining the global coherence of synthesized images. Our discriminator model presents the advantages of both architectures, and we integrate the mechanism of attention \cite{chen2021attention, oktay2018attention}, which allows the discriminator to learn the representations of edges in the images and put emphasis on the selected details. We show that with the new discriminator architecture, we produce more perceptually convincing results than prior works.

\section{Method}
\subsection{Attention enhanced super-resolution GAN Model (A-ESRGAN)}

\begin{figure}
    \centering
    \includegraphics[width=0.48\textwidth]{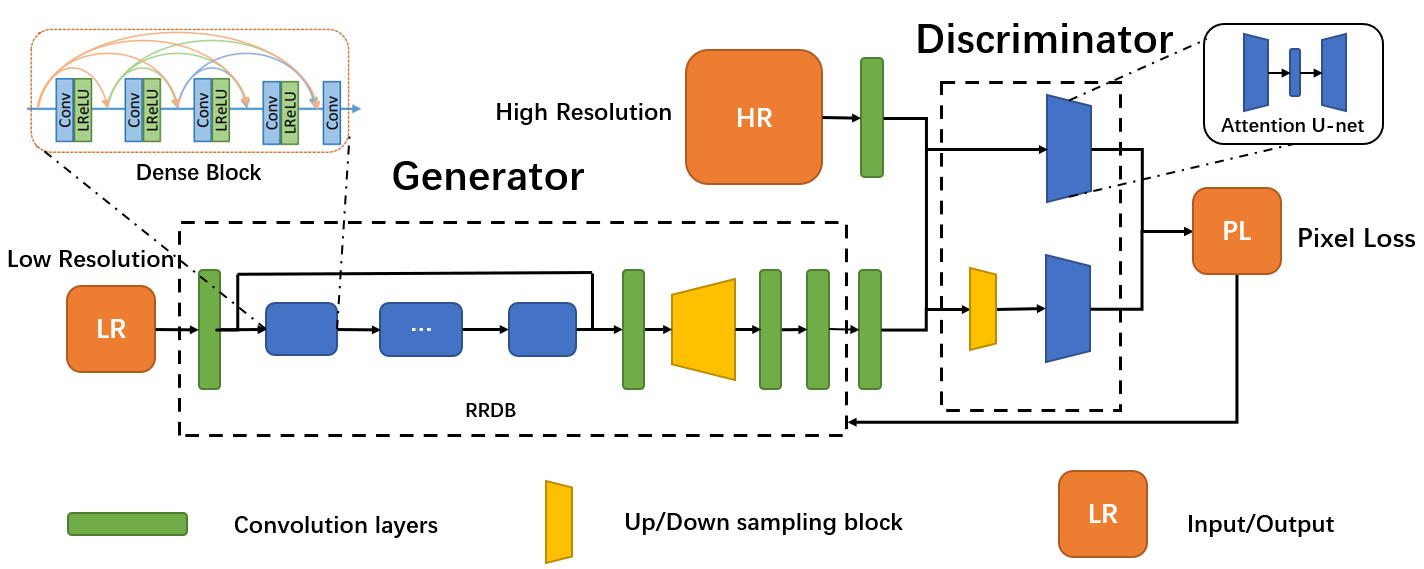}
    \caption{The overall architecture of the A-ESRGAN. The generator of A-ESRGAN is using RRDB, which is adopted from ESRGAN's generator \cite{ESRGAN}.}
    \label{fig:networkArch}
\end{figure}

As shown in \autoref{fig:networkArch}, the proposed network of A-ESRGAN contains a Generator and two separate discriminators, as traditional GAN models. Due to the competition of the two networks, the generator can generate images nearly the same as the real samples.

\par
\textbf{Degradation Model.} We utilized the newly-proposed high-order degradation model \cite{Real-ESRGAN} to synthesize LR images. Compared with traditional first-order degration model, a high-order degration model implements several times the same degration operation and thus better intimate the real-world condition.

\par
\textbf{Generator Architecture.} Stacking residual-in-residual dense blocks (RRDB), shown in \autoref{fig:networkArch}, has shown great performance in SR problems and has been adopted by many SR methods such as  \cite{BSRGAN} and RealESRGAN \cite{Real-ESRGAN}. We also adopted RRDB as our generator.

\begin{figure}
    \centering
    \includegraphics[width=0.48\textwidth]{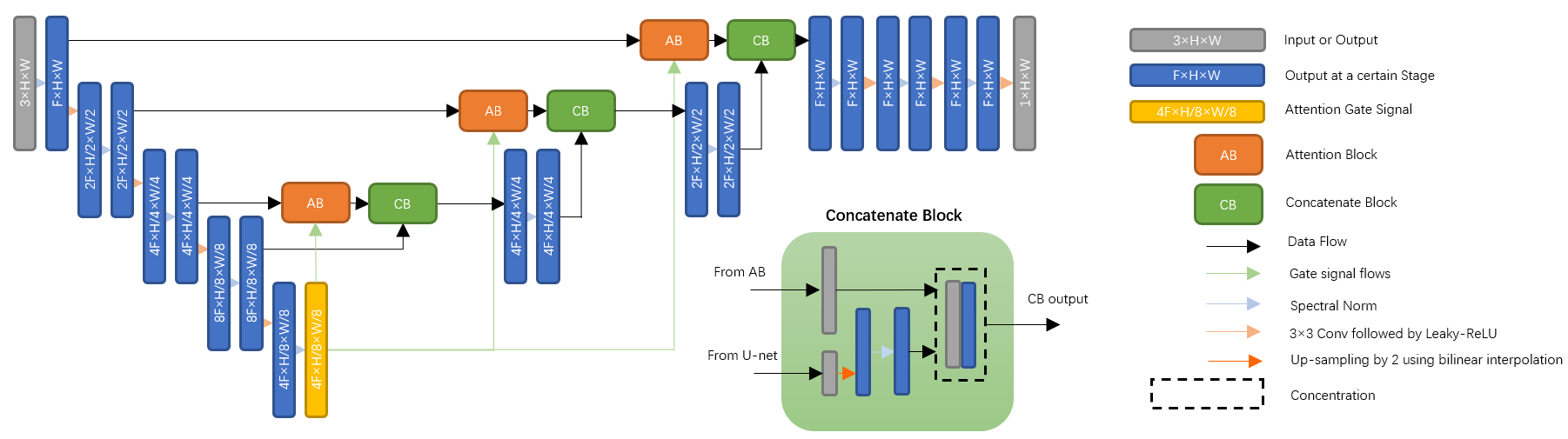}
    \caption{The architecture of a singe attention U-Net Discriminator. F, W, H represents output channel number of the first convolution layer, height of the image and width of the image respectively.}
    \label{fig:UnetArch}
\end{figure}

\begin{figure}
    \centering
    \includegraphics[width=0.48\textwidth]{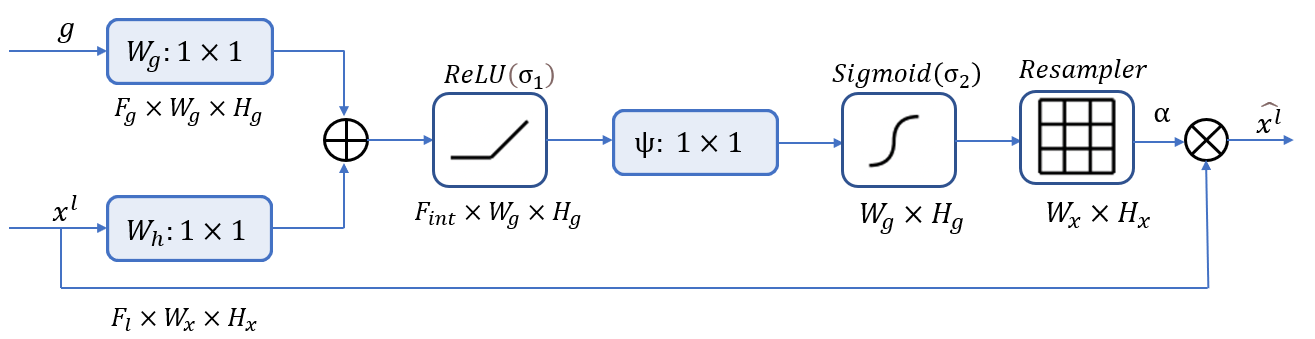}
    \caption{The architecture of the attention block (AB), which is modified from \cite{oktay2018attention}. Here $x^l$ is the input features from the U-Net and $g$ is the gating signal. $F_{int}$ is a super parameter denoting the output channels of the 1 by 1 convolution in the AB. In the AB, $x^l$ is scaled by attention coefficient $\alpha$.}
    \label{fig:AttentionBlock}
\end{figure}

\textbf{Attention U-Net Discriminator.}
Inspired by \cite{Schonfeld_2020_CVPR} and \cite{oktay2018attention}, we propose the attention U-Net discriminator structure, which is shown in \autoref{fig:UnetArch}. It composes a down-sampling encoding module, an up-sampling decoding module and several attention blocks. The detailed structure of the attention block is shown in \autoref{fig:AttentionBlock}. Noted in \cite{oktay2018attention}, the attention gate is used for semantic segmentation of medical images, which is 3D images, so we modified it to use on 2D images. Moreover, following the experience of RealESRGAN \cite{Real-ESRGAN}, we apply spectral normalization regularization \cite{SpectralNorm} to stabilize the training process. 

\par
\textbf{Multi-scale Discriminator} A-ESRGAN adopts a multiple discriminator architecture that has 2 identical attention U-Nets as the discriminator with one discriminator $D_1$ takes an original scale image as input and another discriminator $D_2$ takes a $2\times$ downsampled image as input.

\subsection{The Relativistic Discriminators}
The output of the U-Net discriminator is a $W\times H$ 
matrix and each element denote the likelihood that the pixel it represents is true. To calculate the total loss of one discriminator, we use the sigmoid function to normalize the output and use binary cross-entropy loss to calculate the loss. Assume $C$ is the output matrix, we define $D=\sigma(C)$, $x_r$ is real data and $x_f$ is fake data.

Therefore, uwe define the loss of one discriminator as
\begin{equation}
\begin{split}
    L_D= &\sum_{w=1}^{W}\sum_{h=1}^{H}(-E_{x_r}[\log(D_(x_r,x_f)[w,h])] \\
    &-E_{x_f}[1-\log(D(x_f,x_r)[w,h])])\\    
\end{split}
\end{equation}

Because we have multi-scale discriminators, we will add up the Loss of the discriminators to get the total Loss
{\setlength\abovedisplayskip{0cm}
\setlength\belowdisplayskip{0cm}
\begin{equation}
    L_{Total}=\lambda_1 L_{D_{normal}}+\lambda_2 L_{D_{sampled}}
\end{equation}
    }
    where $\lambda_1$ and $\lambda_2$ are coefficients.
Likely, we can also obtain the generator loss generated by one discriminator
\begin{equation}
\begin{split}
    L_G=&\sum_{w=1}^{W}\sum_{h=1}^{H}(-E_{x_r}[1-\log(D_(x_r,x_f)[w,h])]\\
    &-E_{x_f}[\log(D(x_f,x_r)[w,h])]\\
\end{split}
\end{equation}
    
    Where $x_f$ represents the output of the generator $G(x_i)$

\subsection{Perceptual Loss for Generator}
Apart from the loss obtained from the output of discriminators, we use L1loss and perceptual loss \cite{Perceptual_loss} to better tune the generator.
\par
Thus we obtain the whole loss function for generator
\begin{equation}
    l_G=L_{precep}+\lambda_1 L_{G_{normal}}+\lambda_2 L_{G_{sampled}}+\eta L_1
    \label{lgen}
    \end{equation}
    where $\lambda_1,\lambda_2, \eta$ are coefficients that need to be tuned.
\section{Experiments}
\subsection{Implementing Detail}
To better compare the functionality of multi-scale mechanism, we build 2 A-ESRGAN mofdels: A-ESRGAN-single and A-ESRGAN-multi. The difference is that A-ESRGAN-single features one single attention U-Net discriminator, while A-ESRGAN-multi features multi-scale network, i.e. two identical attention U-Net discriminator operating at different image scale.

We trained with our A-ESRGAN on DIV2K \cite{DIV2K} dataset. 
For better comparison with Real-ESRGAN, we follows the setting of training Real-ESRGAN \cite{Real-ESRGAN} and load the pre-trained Real-ESRNET to the generator of both A-ESRGAN-Single and A-ESRGAN-Multi. The training HR patch size is 256. We train our models with one NVIDIA A100 and three NVIDIA A40 with a total batch size of 48 by using Adam optimizer.
\par 
The A-ESRGAN-Single is trained with a single attention U-Net discriminator for $400K$ iterations under $10^{-4}$ rate. The A-ESRGAN-Multi is trained for $200K$ iterations under $10^{-4}$ learning rate.
\par
For both A-ESRGAN-Single and A-ESRGAN-Multi, the weight for L1loss, perceptual loss and GAN loss are $\{1,1,0.1\}$. The A-ESRGAN-Multi is composed of two discriminators $D_{nromal}$ and $D_{sampled}$, which has the input of 1X and 2X down-sampled images as the input. The weight for GAN loss of $D_{nromal}$ and $D_{sampled}$ is $\{1,1\}$.  The implementation of our model is based on BasicSR \cite{BasicSR}. 

\subsection{Testing Datasets}
In prior works, the synthesized low resolution (LR) images manually degraded from high resolution (HR) are usually used to test the model in blind image super-resolution task.
However, the human simulated degraded images can hardly reflect the low-resolution image coming from degradation in real world, which usually features complicate combinations of different degradation processes. Besides, there is no real dataset which provides real-world LR images. Therefore, we choose to use real-world images, resizing them to 4 times as large as original images and use these as our test dataset.

\par
In this paper, we use the real-world images in the five standard benchmark datasets, Set5 \cite{Set5}, Set14 \cite{Set14}, BSD100 \cite{BSD100}, Sun-Hays80\cite{Sunhays} and Urban100 \cite{Urban100}. 
These five datasets contains images from manifold groups, such as portraits, scenery and buildings. We argue that a good general super resolution model should achieve good performance on the overall 5 datasets.

\subsection{Compared Methods}
We compare the proposed A-ESRGAN-Single and ESRGAN-Multi with several state-of-the-art(SOTA) generative based methods, i.e. ESRGAN \cite{ESRGAN}, RealSR \cite{RealSR}, BSRGAN \cite{BSRGAN}, Real-ESRGAN \cite{Real-ESRGAN}. Note that the architecture of the generators of ESRGAN, BSRGAN and Real-ESRGAN are the same as us, which can help verfiy the effectiveness of our designed discriminator.

\subsection{Experiment Results}
\begin{figure}
    \centering
    \includegraphics[width=0.5\textwidth]{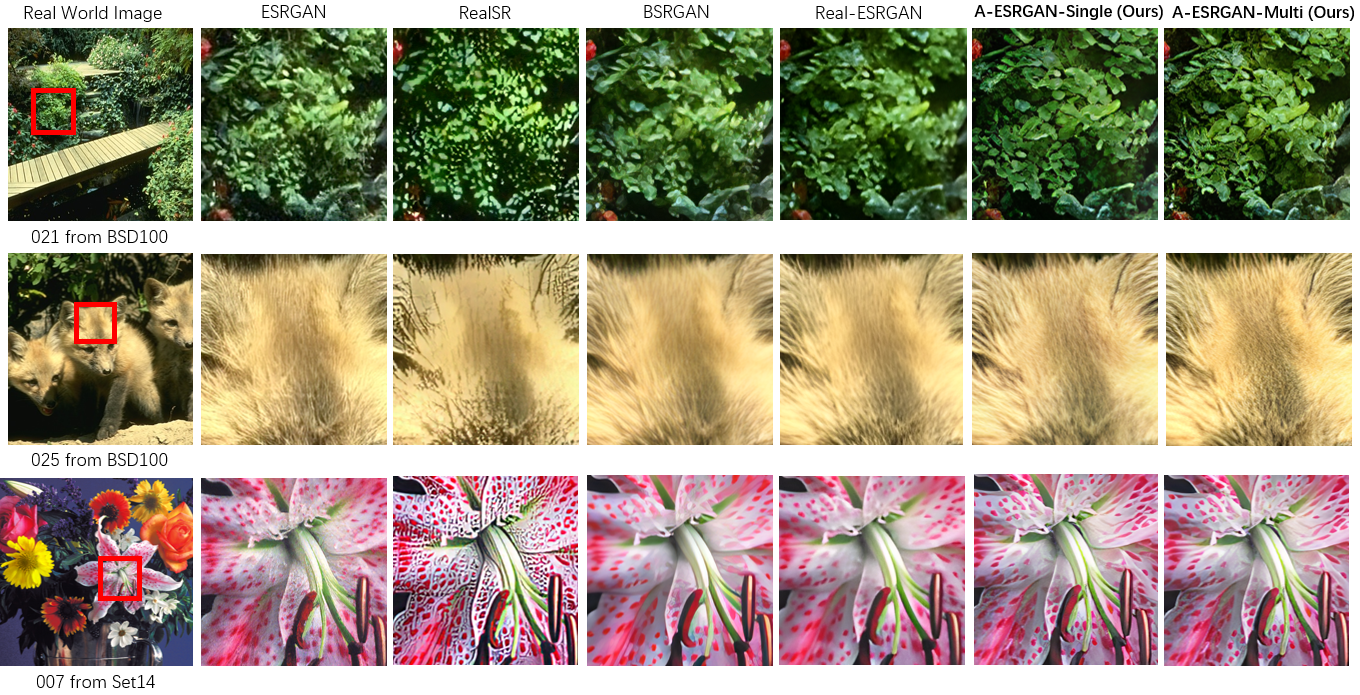}
    \caption{Visual comparison of our method with other $\times$4 super resolution methods. Zoom in for the best view.}
    \label{fig:ViusalCompare}
\end{figure}

Since there is no ground-truth for the real world images of the dataset, so we adopt the no-reference image quality assessment metrics NIQE \cite{NIQE} for quantitative evaluation. NIQE indicates the perceptual quality of the image.  A lower NIQE value indicates better perceptual quality. As can be seen from the \autoref{table:Metrics}, our method outperforms most of the SOTA methods in NIQE metrics. From visual comparison (some examples are shown in \autoref{fig:ViusalCompare}), we observe our methods can recover sharper edges and restore better texture details. 

\begin{table*}[htbp]
\resizebox{\textwidth}{!}{
\begin{tabular}{lccccccc}
\toprule
\rowcolor[HTML]{EFEFEF} 
NIQE & Bicubic & ESRGAN & BSRGAN & RealESRGAN & RealSR & \textbf{A-ESRGAN-Single(Ours)} & \textbf{A-ESRGAN-Multi(Ours)} \\ 
\midrule
\rowcolor[HTML]{FFFFFF} 
\cellcolor[HTML]{EFEFEF}Set5       & 7.8524 & 5.6712 & 4.5806 & 4.8629 & \textcolor{red}{3.5064} & 3.9125          & \textcolor{blue}{3.840}          \\
\rowcolor[HTML]{FFFFFF} 
\cellcolor[HTML]{EFEFEF}Set14      & 7.5593 & 5.0363 & 4.4096 & 4.4978 & 3.5413          & \textcolor{red}{3.4983}          & \textcolor{blue}{3.5168} \\
\rowcolor[HTML]{FFFFFF} 
\cellcolor[HTML]{EFEFEF}BSD100     & 7.3413 & \textcolor{red}{3.1544} & 3.8172 & 3.9826 & 3.6916          & 3.2948 &   \textcolor{blue}{3.2474}              \\
\rowcolor[HTML]{FFFFFF} 
\cellcolor[HTML]{EFEFEF}Sun-Hays80 & 7.6496 & 3.6639 & 3.5609 & 2.9540 & 3.3109          & \textcolor{blue}{2.6664} &        \textcolor{red}{2.5908}         \\
\rowcolor[HTML]{FFFFFF} 
\cellcolor[HTML]{EFEFEF}Urban100   & 7.1089 & \textcolor{red}{3.1074} & 4.1996 & 4.0950 & 3.929           & 3.4728 &         \textcolor{blue}{3.3993}       \\ \bottomrule
\end{tabular}
}

\caption{The NIQE results of different methods on Set5, Set14, BSD100, Sun Hays80 and Urban 100 (The lower, the better). The best and second best results are high lighted in red and blue, respectively. \label{table:Metrics} }
\end{table*}

\subsection{Attention Map Analysis}

\begin{figure}[htbp]
\centering
\includegraphics[width=0.5\textwidth]{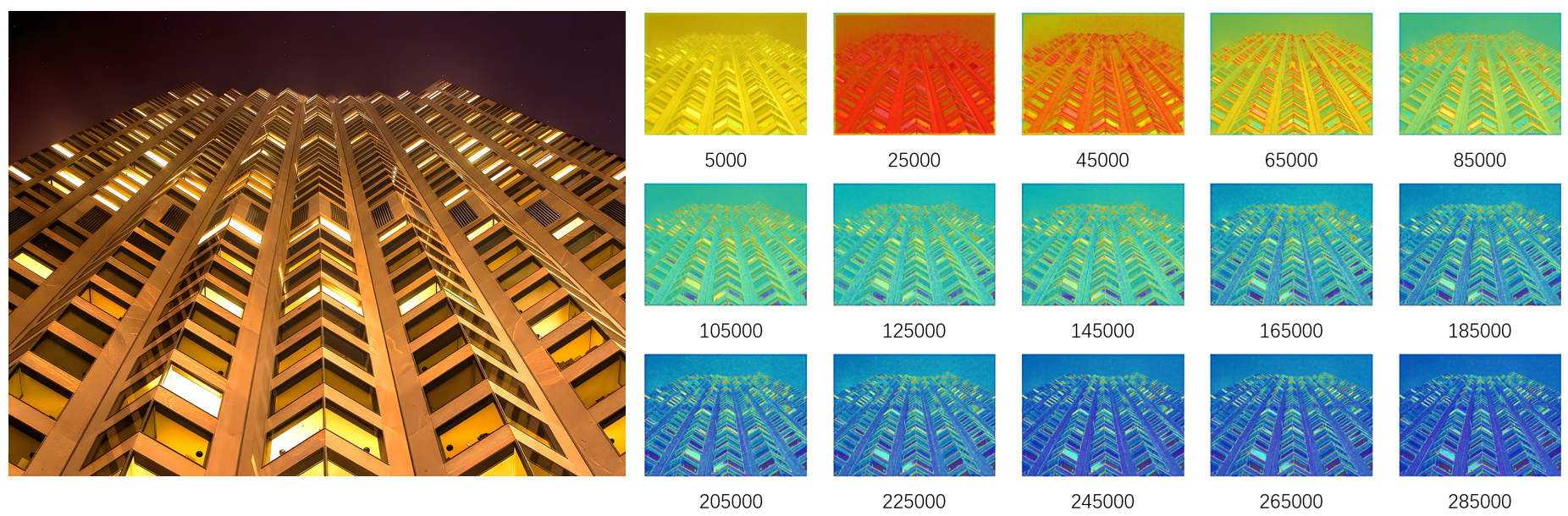}
\caption{ The figure shows the weight in the third attention layer across the training process from iteration 5000 to 285000 at an interval of 20000. The example image is picked from Urban100 \cite{Urban100}. It clearly shows at first the attention is uniformly distributed. Then the attention is gradually updated and begins to focus on the edges. Zoom in for the best view.\label{fig:AttentionViusal}}
\end{figure}

\begin{figure}[htbp]
\centering
\includegraphics[width=0.5\textwidth]{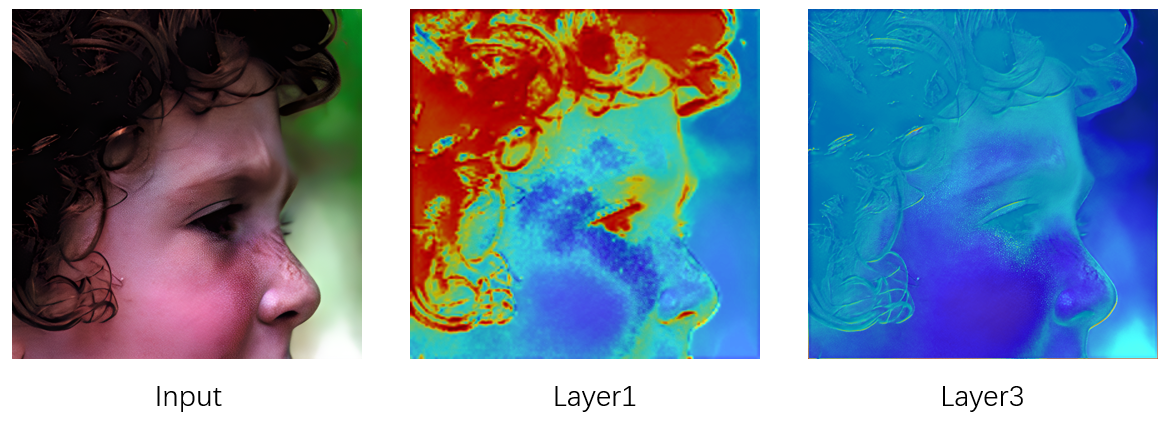}
\caption{ The figure shows the weight of the first and third attention layer at iteration 200000. The example image is picked from Set5 \cite{Set5}. The example shows lower level attention(first) would learn coarse-grained color changes (patches) while upper level attention(third) learn fine-grained color changes (dots and lines). The layers are resized for better view.\label{fig:DifferentLayerAttention}}
\end{figure}

To verify the effectiveness of attention gate in our discriminator, We visualize the attention weights in the attention layer from test images during our training process. The example is shown in \autoref{fig:AttentionViusal}. Initially, the attention weights are uniformly distributed in all locations of the images. As the training process goes on, we can observe that the attention weight is gradually updated and begin to focus on "particular regions", which are the edges where color changes abruptly. Meanwhile, by visualizing attention map at different layers, we argue that different attention layers recognize the images at different granularity. As shown in \autoref{fig:DifferentLayerAttention}, the lower attention layers are coarse-grained give rough edges of the patches while the upper attention layers are fine-grained and focus on details such as lines and dots.

\subsection{Disctiminator output analysis}

We study the output image generated by the two attention U-Net disctiminators and propose that the two discriminators play different roles in identifying the properties of the images. The normal discriminator, which is also used in the single version, emphasizes more on lines. In contrast, the input downsampled input images with blurred edges force the other discriminator to focus more on larger patches. \par
\begin{figure}[htbp]
\centering
\includegraphics[width=0.5\textwidth]{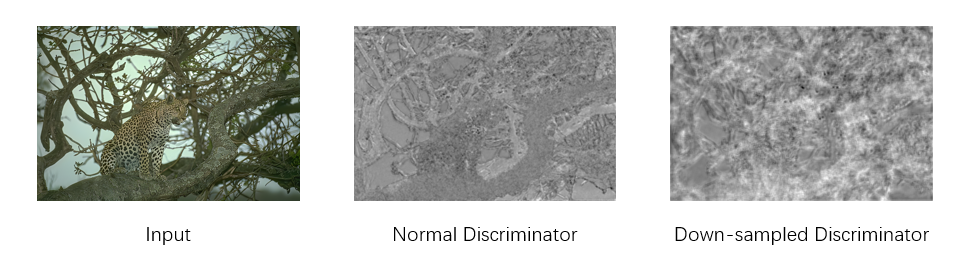}
\caption{ The figure shows Unet output of the two discriminators. The example image is picked from BSD100 \cite{BSD100}. The example shows the normal discriminator(first) would focus on lines in the image while the discriminator that parse the downsampled input will focus on patches. The brighter a pixel is the more likely it is going to be a real picture.
The outputs are resized for better view.\label{fig:DifferentUnetOutput}}
\end{figure}

As shown in \autoref{fig:DifferentUnetOutput}, the output image of the normal discriminator judges the edges while the the dowsampled discriminator judges thicker blocks, such as textures on the branches of the tree.

\subsection{Ablation Study}
\par
\textbf{Effect of attention U-Net discriminator.} 
The key factor of A-ESRGAN surpassing the existing models is our designed attention U-Net discriminator. In the ablation study, we compare the results of Real-ESRGAN model and A-ESRGAN-Single model. The only difference of these two network is that Real-ESRGAN uses a plain U-Net as discriminator, while A-ESRGAN applies an attention U-Net discriminator.
\par
As shown in \autoref{table:Metrics}, A-ESRGAN-Single achieves better NIQE in all tested datasets. By taking a close look at the result, we could find since plain U-Net uniformly gives weight to each pixel, it can't distinguish between the subject area  and background of images. However, as shown in Section 4.5, the attention U-Net is able to put more efforts on the edges than ordinary pixels.
\par
\begin{figure}[htbp]
\centering
\subfloat[\label{fig:AttentionUnetAblationSharper}]{\includegraphics[width=0.45\textwidth]{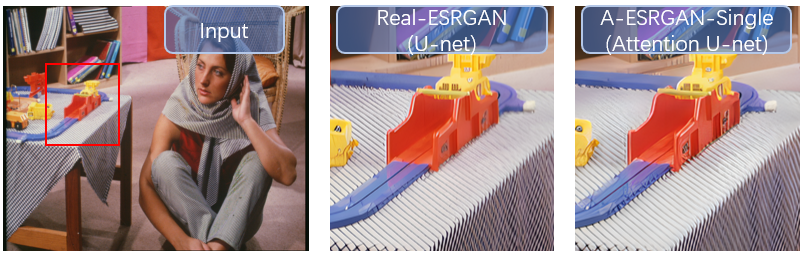}}
\quad
\subfloat[\label{fig:AttentionUnetAblationDistortion}]{\includegraphics[width=0.45\textwidth]{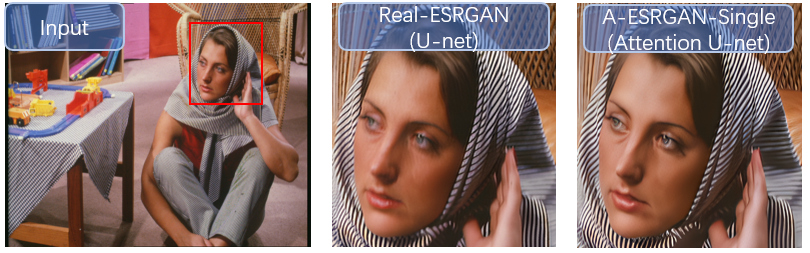}}
\caption{ Ablation on the discriminator design. Zoom in for the best view.}
\end{figure}

We believe this will bring at least two benefits. First, the result image will give sharper and clearer details as shown in \autoref{fig:AttentionUnetAblationSharper}. Second, when up-sampling process is based on the main edges of the image, there will be less probability of distortion (like shown in \autoref{fig:AttentionUnetAblationDistortion}). 

\textbf{Effect of multi-scale discriminator.} 
The multi-scale discriminator  enable our model to focus on not only the edges but also on more detailed parts such as textures. In the ablation study, we compare the result of the A-ESRGAN-single and the A-ESRGAN-multi. The latter has the same generator as the former while it possesses two discriminators, which are a normal one and a downsampled one.
\par
\begin{figure}[htbp]
\centering
\subfloat[\label{fig:multiAblation1}]{\includegraphics[width=0.45\textwidth]{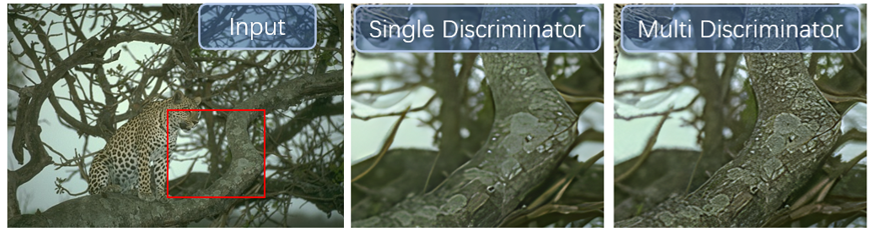}}
\quad
\subfloat[\label{fig:multiAblation2}]{\includegraphics[width=0.45\textwidth]{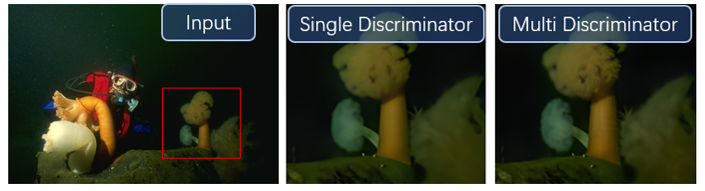}}
\caption{ Ablation on the multi-scale design. Zoom in for the best view.}
\label{fig:multiAblation}
\end{figure}

As shown in \autoref{table:Metrics}, the A-ESRGAN-multi surpasses the performance of A-ESRGAN-single in all dataset except Set14. By analyzing the output images of the two models, we conclude that the A-ESRGAN-multi does much better on showing the texture of items than A-ESRGAN-single. Like the images shown in \autoref{fig:multiAblation}, the A-ESRGAN-single poorly performs on rebuilding the texture of the branches and the sea creature. In contrast, because the downsampled discriminator focus on patches, it can rebuild the texture as well as give shaper edge details.

\section{Conclusions}
In this paper, a multi-scale attention U-Net discriminator is proposed to train a deep blind super-resolution model. Based on the new discriminator, we trained a deep blind super-resolution model and compared it with other SOTA generative methods by directly upscaling real images in 5 benchmark datasets. Our model outperforms most of them in both NIQE metrics and visual performance. By systematically analyzing how the attention coefficient changes across time and space during the training process, we give a convincing interpretation of how the attention layer and multi-scale mechanism contribute to the progress in SR problems. We believe that other super-resolution models can benefit from our work.


\bibliographystyle{IEEEbib}
\bibliography{chapters/reference}

\end{document}